\documentclass[conference]{IEEEtran}

\usepackage{amsmath,amssymb}
\usepackage{booktabs}
\usepackage{microtype}
\usepackage{algorithmic}
\usepackage[linesnumbered,ruled,vlined]{algorithm2e}
\usepackage{subcaption}
\usepackage{url}

\usepackage{bm}
\usepackage{mathtools}
\usepackage{graphicx}

\newcommand{\method}{MACPruning\xspace}
\usepackage{xcolor}
\usepackage{fancybox}

\definecolor{benign}{HTML}{98c47a}
\definecolor{boxcolor}{HTML}{f2f2f2}%
\definecolor{theo_color}{HTML}{3f2e19}%
\definecolor{mes_color}{HTML}{a83c23}%
\definecolor{bleudefrance}{rgb}{0.19, 0.55, 0.91}
\usepackage{float}

\usepackage{siunitx}
\usepackage{cite}
\usepackage{multirow}
\usepackage{multicol}

\newboolean{showcomments}
\setboolean{showcomments}{False}
\newcommand{\fei}[1]{{\ifthenelse{\boolean{showcomments}} {\color{red}\emph{[Fei: #1]}}{}}}
\newcommand{\ding}[1]{{\ifthenelse{\boolean{showcomments}} {\color{red}\emph{[Ding: #1]}}{}}}
\newcommand{\cheng}[1]{{\ifthenelse{\boolean{showcomments}} {\color{teal}\emph{[Cheng: #1]}}{}}}
\newcommand{\ruyi}[1]{{\ifthenelse{\boolean{showcomments}} {\color{green}\emph{[Ruyi: #1]}}{}}}
\newcommand{\update}[1]{{\ifthenelse{\boolean{showcomments}} {\color{teal}{#1}}{#1}}}

\begin{document}

\title{MACPruning: Dynamic Operation Pruning to Mitigate Side-Channel DNN Model Extraction}

\author{
\IEEEauthorblockN{Ruyi Ding, Cheng Gongye, Davis Ranney, Aidong Adam Ding, Yunsi Fei}
\IEEEauthorblockA{Northeastern University\\
  \{ding.ruy, gongye.c, ranney.d, a.ding, y.fei\}@northeastern.edu}
}

\maketitle

\begin{abstract}
As deep learning gains popularity, edge IoT devices have seen proliferating deployment of pre-trained Deep Neural Network (DNN) models. 
These DNNs represent valuable intellectual property and face significant confidentiality threats from side-channel analysis (SCA), particularly non-invasive Differential Electromagnetic (EM) Analysis (DEMA), which retrieves individual model parameters from EM traces collected during model inference. 
Traditional SCA mitigation methods, such as masking and shuffling, can still be applied to DNN inference, but will %
incur significant performance degradation due to the large volume of operations and parameters.
Based on the insight that DNN models have high redundancy and are robust to input variation, we introduce MACPruning, a novel lightweight defense against DEMA-based parameter extraction attacks, exploiting specific characteristics of DNN execution.
The design principle of MACPruning is to randomly deactivate input pixels and prune the operations (typically multiply-accumulate--MAC) on those pixels. 
The technique removes certain leakages and overall redistributes weight-dependent EM leakages temporally, and thus effectively mitigates DEMA. 
To maintain DNN performance, we propose an importance-aware pixel map that preserves critical input pixels, keeping randomness in the defense while minimizing its impact on DNN performance due to operation pruning. 
We conduct a comprehensive security analysis of MACPruning on various datasets for DNNs on edge devices. Our evaluations demonstrate that MACPruning effectively reduces EM leakages with minimal impact on the model accuracy and negligible computational overhead.
\end{abstract}

\section{Introduction}
Side-channel analysis (SCA) is a security exploit that focuses on inferring secret information from side-channel leakage collected during the execution of software or hardware systems.
Commonly-used side-channel leakages include power consumption~\cite{wei2018know, xiang2020open, randolph2020power}, electromagnetic (EM) emanations~\cite{kasper2009side, longo2015soc}, and execution time~\cite{jiang2017novel, zhang2016cloudradar}.
SCA  has primarily targeted %
cryptographic systems for recovering the secret keys or sensitive messages. 
Recently, with the wide deployment of Deep Neural Networks (DNNs), the pre-trained DNN models become a lucrative target of SCA~\cite{batina2019csi, xiang2020open,  ding2023emshepherd, wei2020leaky}.
Particularly for intelligent edge devices, such as FPGAs and Microcontroller Units (MCUs), physical access or proximity makes it easy to collect side-channel power and EM traces~\cite{batina2019csi, gongye2023side}.
Valuable intellectual property (IP) of DNNs includes network structures, hyperparameters, and parameters such as weights and biases, which are all susceptible to SCA. Such attacks not only compromise the model confidentiality and infringes IP, but also facilitates other serious attacks including integrity breaches~\cite{madry2017towards, tramer2020adaptive, rakin2019bit} and model inversions ~\cite{rakin2022deepsteal, dmitrenko2018dnn}.

Existing defense mechanisms against SCA on cryptographic systems, such as masking and shuffling, can be applied to DNN models too, but often incur a a large execution overhead.
Modern ciphers involve simple operations and a small amount of secrets, and thereby are not computation-intensive.  
In a contrast, DNN models are composed by high volume of parameters and involve enormous operations, making the prior mitigation methods computational prohibitive. %
For instance, %
masking an embedded DNN results in a $127\%$ increase in the number of cycles compared to the unmasked DNNs~\cite{dubey2020maskednet}.
Hiding techniques such as operation shufflingalso result in substantial computation and storage overheads, with a reported latency increase of $18\%$~\cite{brosch2022counteract}.
We realize that previous defenses have not considered the distinct characteristics of DNN models, such as their inherent tolerance of input variations. 
To bridge the gap between general side-channel countermeasures and the unique features of DNN models, we introduce \method.
Our approach leverages the robustness of DNN models to input variations to protect the model confidentiality with high efficiency and minimal DNN performance degradation.

\fei{what do you want to say for these prior masking methods? they are not for side-channel mitigation. have to point it out}
\ruyi{I want to provide more background of input masking of DNN, to show our motivation of preventing EM side channel. Revise here.}
DNNs have been demonstrated to be robust to missing features of the model inputs, a property that has been utilized in model training~\cite{he2022masked} and inference~\cite{gao2017deepcloak} to improve their generalizability and adversarial robustness.
Leveraging such tolerance of incomplete or partially occluded inputs~\cite{zhang2013occlusion}, our proposed \method, for the first time, applies dynamic input dropping to defend DNN execution against SCA-based model extraction attacks on edge devices.
\method proposes a random pixel activation map (called \texttt{RPAM}) to be applied on the model input, where a subset of pixels is deactivated and the subsequent operations on them are pruned at run-time, effectively protecting the model parameters against DEMAs, which rely heavily on the sequential execution order of leaky operations and precise leaky locations. 

Compared with previous SCA defense methods, \method is lightweight and achieves high efficiency, as our defense is based on random operation-skipping, instead of shuffling. %
However, \method may degrade the DNN performance (classification accuracy) as it reduces the amount of input information during inference.
To minimize the performance degradation due to the loss of information from deactivated pixels, we introduce an importance-aware pixel activation map (\texttt{IaPAM}). 
Specifically, critical pixels are identified and excluded from deactivation, while other less important pixels are randomly dropped at certain prescribed rate. 
Preserving critical pixels ensures the DNN inference performance, and the randomness of deactivating other pixels offers SCA mitigation. 
The IaPAM is trained using a modified loss function that balances the trade-off between defense efficacy and DNN performance.

The major contributions of this paper are the following:
\begin{itemize}
    \item \method, a novel and lightweight defense mechanism, is proposed to protect DNN models against SCA-assisted model extraction attacks. The advantage of our method over traditional SCA countermeasures is that characteristics of DNN models are utilized to achieve both security and efficiency.
    \item A pixel importance aware strategy is introduced and incorporated in \method to preserve performance-critical pixels, so as to strike a balance between security and model performance.
    \item Comprehensive theoretical analysis and rigorous evaluations on real EM measurements from an actual microcontroller are conducted. 
    Specifically, we show that the proposed \method with \texttt{RPAM} and \texttt{IaPAM} offers adequate model confidentiality protection with a small overhead (less than $3\%$).
\end{itemize}

\section{Related Work}

\subsection{EM Side-channel Analysis}
EM Side-channel Analysis, including simple EM analysis and Differential EM Analysis (DEMA), exploits EM emissions to extract sensitive information, including cryptographic keys, as well as DNN model hyperparameters and parameters.
DEMA specifically utilizes the data-dependency of EM signals and employs statistical analysis over a set of EM traces to infer the secrets.
For a Device Under Test (DUT), the adversary builds an EM leakage model for key modules of the system implementation, e.g., a register holding secret-dependent intermediate values, the result of certain operations.
The model predicts EM emissions based on hypothetical secret values, and these predictions are then correlated with actual EM measurements.  
Correct secret guesses will yield the highest correlations. 
A common model used in DEMA against DNN parameters on MCUs is the Hamming Weight (HW) model, which counts the number of `1' bits in a variable's binary representation~\cite{batina2019csi}. To improve the efficiency, often times a range of points of interests are selected from the traces to correlate with the EM predictions, based on analysis of execution of the sequence of operations. 

\subsection{Defense against Side-channel Analysis}
To counter such attacks, several countermeasure principles, previously applied to ciphers, have been adapted for DNN model protection. Masking has been applied both to hardware implementations ~\cite{dubey2020maskednet} and software platforms~\cite{popets2022} to thwart parameter extraction attacks. Hiding techniques, such as operation shuffling, have also been implemented on DNN solutions~\cite{dubey2022guarding,brosch2022counteract}. Maji et al.\cite{maji2022threshold} employed threshold implementation, while Hashemi et al.\cite{hashemi2022hwgn} utilized garbled circuits for DNN protection. All the prior approaches, although effective in protecting DNN models against parameter extraction attacks, bear significant implementation costs by dealing with a lot of operations directly, neglecting special DNN features.
\emph{To the best of our knowledge, our proposed approach is the first defense that diverges from the traditional SCA mitigation techniques and leverages DNN characteristics for efficient SCA protections.
}

\section{Our Approach: \method} \label{sec: methodology}
This section outlines the methodology of our approach \method. We first present the threat model and summarize the characteristics of DNNs under SCA. We then illustrate how to leverage these characteristics for an efficient and effective SCA defense approach. Finally, we describe how to train the importance-aware pixel activation map in detail. 

\begin{figure}[t]
  \centering\includegraphics[width=\linewidth]{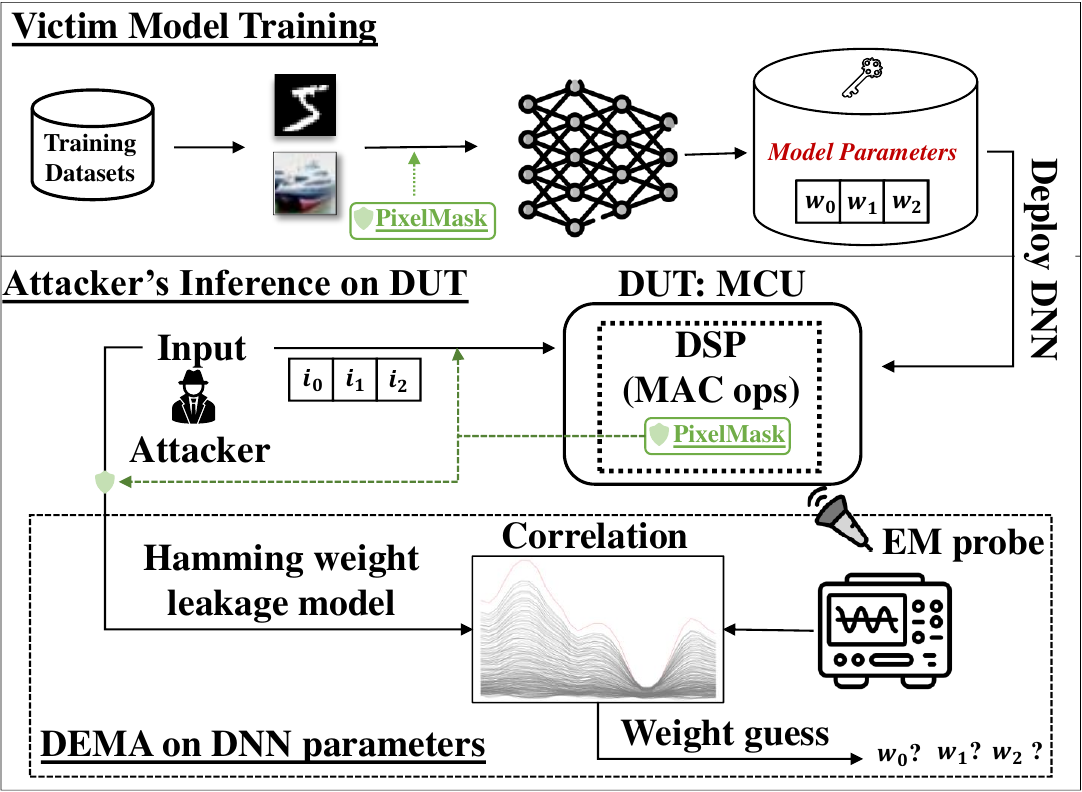}
  \caption{Illustration of the threat model. The victim model is deployed on an edge device MCU. The attacker is able to conduct DEMA to retrieve {\color{red!50}secret parameters}. Our proposed protection {\color{benign!70} \method} is deployed during the DNN training and inference phase to hide side-channel leakage.}
  \label{fig: threat_model}
\end{figure}

\begin{figure*}[t]
    \centering
    \includegraphics[width=\linewidth]{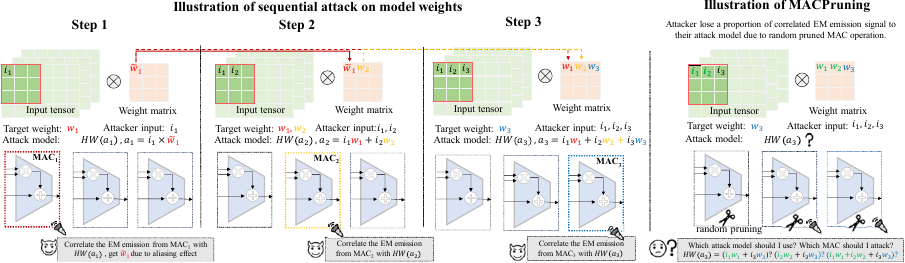}
    \caption{\textbf{DEMA attack procedure on MCU against edge DNNsand \method defense rationale:} the attacker leverages DEMA attack by correlating the EM leakage signals with the cumulative values in a sequential manner, they must attack the target weights $w_1$, $w_2$, $w_3$ in order. Note that when attack $w_1$, they can only get possible values of $w_1$ due to aliasing, denoted as $\tilde{w}_1$. The real value of $w_1$ can be confirmed in Step 2. By randomly deactivate the input pixel, \method drops the corresponding MAC operations so that the percentage of vulnerable EM signal reduced.}
    \label{fig: attack process}
\end{figure*}
\subsection{Threat Model} \label{sec: threat model}

In line with the common threat model for EM side-channel analysis, we delve into scenarios where an adversary attempts to extract secret model parameters through DEMA. We provide the setup of victim implementation and define the adversary's knowledge and capabilities. Fig.~\ref{fig: threat_model} presents the threat model.

\subsubsection{Victim}
In Figure~\ref{fig: threat_model}, the victim DNN model is deployed on the MCU of an edge device, equipped with standard inference firmware, such as TensorFlow Lite Micro\footnote{\url{https://www.tensorflow.org/lite/microcontrollers}}. The victim model is trained \textit{a prior} with a private dataset, and the model weights and biases are securely loaded in the MCU program memory. The model parameters are all fully quantized 8-bit signed integers, adhering to the TensorFlow Lite Micro specifications. Notably, a typical commodity MCU has an optional DSP for performing MAC operations.

\subsubsection{Attacker's Knowledge}
We posit an informed attacker, who has knowledge of the model hyperparameters, including the structure and activation functions of the victim model. Additionally, they know the DUT implementation details, including the sequence of MAC operations. However, they do not know the model parameters (weights and biases), and aim to retrieve them to steal the IP or facilitate other white-box attacks.

\subsubsection{Attacker's Capability}
The attacker has physical access to the DUT or is in proximity, and can capture EM emissions of model inference in traces. Furthermore, he controls inputs to the victim model and monitors the inference output. Techniques such as direct program memory dumping, bus snooping, fault injection, and learning-based model extraction attacks fall outside the scope of this work.

\subsection{DEMA Attack Procedure on MCU}
In Fig.~\ref{fig: attack process}, we illustrate the process of a DEMA attack. Specifically, the chosen leakage model is the Hamming weight of cumulative values of user inputs and model weights (i.e., MAC results), instead of individual products of a weight and an input due to the strong aliasing effects of such operations~\cite{gongye2023side}.
As shown in the example, the attacker follows the order of operations and targets $w_0$, $w_1$, and $w_2$ sequentially. The proposed \method obfuscates the leaky signals on the EM traces by randomly deactivating pixels according to a predefined pixel map. 
\method significantly reduces the usable signals, with the reduction occurring at an exponential rate under normal attack conditions.
Potential adaptive attacks, utilizing multiple leaky points, will be discussed in Section~\ref{sec: adaptive}.

\subsection{Key Observations} \label{sec: observations}
In this section, we describe the distinct characteristics of quantized DNN model inference on MCUs under DEMA attacks, which our defense strategy  takes into consideration.

\begin{figure}[H]

  \noindent\hspace*{-\fboxrule}
  \begin{Sbox}
    \begin{minipage}{0.95\linewidth}
      \textbf{Observation 1 (O1).}
      DEMA targeting DNN parameters on MCUs utilizes the HW model over accumulated MAC outputs.
    \end{minipage}
  \end{Sbox}

  \fcolorbox{black}{boxcolor}{\TheSbox}
\end{figure}

For DNN inference on MCUs, the in-order execution of instructions renders them particularly vulnerable to DEMA attacks.
With time points for operations that process secret parameters pinpointed on EM traces, adversaries can apply DEMA across the set of traces to recover the parameters with an effective leakage model that depicts the dependence of the side-channel leakage on the secret.

In cryptographic systems such as AES, DEMA employs the HW model on non-linear operations like the Rijndael Substitution.
The non-linearity yields strong distinguishability for the correct secret value.
However, operations in DNN models are predominantly linear, and the HW model on them does not work for SCA anymore. For instance, a common multiplication operation ($i{\cdot}w$, where $w$ is the secret parameter and $i$ is the known and controllable input) demonstrates a significant aliasing effect in the HW of the result, especially when $w$ is a power of 2. This effect, affecting the majority of weight values (191 out of 256)~\cite{gongye2023side}, limits the usefulness of the HW model over individual operations in DEMAs on DNNs.

A more viable leakage model for DEMA on DNNs focuses on the cumulative results of multiplications.
For instance, we first attack $w_1$ with the HW of $i_1{\cdot}w_1$ but get a set of candidates of $w_1$ due to aliasing. Then, the sum $i_1{\cdot}w_1 + i_2{\cdot}w_2$ can be used to infer weights $w_1$ and $w_2$ without the aliasing effect, owing to the non-linear addition operation. Subsequent operations, such as adding $i_3{\cdot }w_3$ to this sum, would reveal $w_3$ at the next leaky time point, and so on.
Therefore, a leakage model for effective DEMA targeting DNN parameters on MCUs, shifts from individual operations to the aggregate result of a series of operations.
Identifying and analyzing such sequences of accumulative operations becomes essential for a successful DEMA attack to retrieve secret parameters.

\begin{figure}[H]

  \noindent\hspace*{-\fboxrule}
  \begin{Sbox}
    \begin{minipage}{0.95\linewidth}
      \textbf{Observation 2 (O2).}
      Safeguarding the first layer of DNNs is pivotal for effective and efficient SCA defense.
    \end{minipage}
  \end{Sbox}
  \fcolorbox{black}{boxcolor}{\TheSbox}

\end{figure}

DNN layers operate in a feed-forward fashion during inference, i.e., each layer's outputs feed to the next layer as inputs. With such data dependency, parameter extraction typically starts with the first layer and progressively proceeds to the subsequent layers.
In our threat model, this layer-by-layer strategy is crucial for an attacker aiming to completely extract the DNN model.
When the first layer is safeguarded, an attacker is hindered from learning the first layer output, which is required for attacking the following layers iteratively.
Hence our defense mechanism against DEMA attacks only needs to focus on protecting DNN's first layer.

\begin{figure}[H]

  \noindent\hspace*{-\fboxrule}
  \begin{Sbox}
    \begin{minipage}{0.95\linewidth}
      \textbf{Observation 3 (O3).}
      DNNs possess the inherent tolerance of input variations, a feature that can be leveraged for effective defenses against EM side-channel attacks.
    \end{minipage}
  \end{Sbox}

  \fcolorbox{black}{boxcolor}{\TheSbox}
\end{figure}

\begin{figure}[t]
  \centering
  \includegraphics[width=\linewidth]{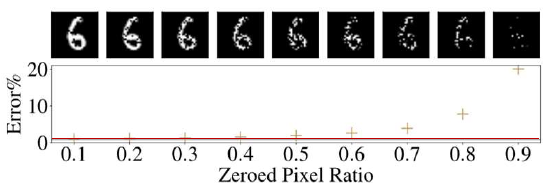}
  \caption{DNN classification accuracy with the pixel zeroization ratio from  $10\%$ to $90\%$ for an MNIST sample.}
  \label{fig: random mnist}
\end{figure}

For image classification tasks, the human brain demonstrates remarkable robustness--capable of recognizing objects even when only parts of them are visible.
DNNs, inspired by human neural processing, retain similar robustness and are able to withstand input variations, specifically pixel losses.
We conduct a preliminary experiment. For samples in the MNIST dataset, the pixel value is only zero (black) or one (white). We pick a sample, randomly set some pixels to zero (black), and evaluate the classification errors of a corresponding DNN.
Fig.~\ref{fig: random mnist} shows the results for digit `6': the sample remains visually identifiable even as the ratio of zeroed-out pixels increases from $10\%$ to $90\%$; while the DNN's error rate escalates with the loss of pixels, it largely maintains its classification accuracy even when the zeroization ratio reaches $50\%$ (accuracy of the baseline model is marked as the bottom {\color{red} red} line).

These phenomena motivate our distinctive defense strategy for DNNs, diverging from the traditional side-channel countermeasures used in protecting ciphers against DEMA.
For example, the shuffling method does not reduce the number of operations, and merely moves them around to break alignment of leaky signals on the traces, required by differential EM analysis. 
In contrast, our \method focuses on the input and the first layer of the DNN, based on the three prior observations, leveraging the inherent input variation tolerance of DNNs and the feed-forward layer-wise data dependency.
This method effectively mitigates DEMA-based model extraction with negligible or even negative execution overhead while maintaining the model accuracy.

\subsection{Defense Overview} \label{sec: defense-overview}

Based on observations \textbf{O1} and \textbf{O2}, we aim to change the sequence of MAC operations in a DNN's first layer to counteract EM DNN parameter extraction attacks. 
Observation \textbf{O3} indicates that by deactivating some input pixels and dropping the corresponding MAC operations, DNN performance might not be affected much.
Thus, altering a DNN's input to disrupt MAC operations is a viable defense strategy. 
We introduce our first defense against DEMA--\textit{Random Pixel Activation Map} (\texttt{RPAM}), as detailed in Section~\ref{sec:rpa}.
\texttt{RPAM} selectively activates a subset of pixels in each inference. To equip the model with such robustness, we also consider in-complete inputs during the training phase of embedded DNNs, as shown in Fig.~\ref{fig: threat_model}. 
The embedded DNN is fine-tuned on images whose pixels are randomly deactivated with $1-p$ during the training phase. 
\ding{This sounds like that you randomly select some inputs to deactivate. Then these selection is fixed. The images with these fixed deactivation is used to fine-tune the DNN. Is this true? I thought that the point of fine-tuning is that the DNN is robust to inputs randomly dropping $1-p$ proportion, similar to a random drop-out network. It is not supposed to be specific to a particular $p$ activated inputs pattern. If so, we should say that ''... DNN is fine-tuned on images with $1-p$ proportion of randomly deactivated inputs." }
\ruyi{It is the second case, the DNN is fine-tuned with random inputs }
We defined the \textit{pixel activation ratio as $p$.} 
During inference, a random \texttt{RPAM} with the rate $p$ will also be applied to the inputs. 
This random skipping of MAC operations eliminates some signals on the EM traces and moves later signals to earlier time points, exponentially increasing the complexity of trace analysis over time.
Moreover, if a pixel is inactive, a group of MAC operations are bypassed, thereby reducing the computation and leading to faster DNN inference.
While activating pixels randomly is effective for simpler datasets (e.g., MNIST), for more complex datasets, an \textit{Importance-aware Pixel Activation Map} (\texttt{IaPAM}) can preserve the inference accuracy better, as detailed in Section~\ref{sec:ipam}.

\subsection{Random Pixel Activation Map} \label{sec:rpa}

\texttt{RPAM} enhances the robustness of DNN models against side-channel model extraction attacks by redistributing the leakage signals in the time domain. 
Consider two consecutive MAC operations, and the accumulator output is $s$. The first MAC operation is $s = s + i_1w_1$, and the second $s = s + i_2w_2$. As per \textbf{O1}, attackers aim to deduce $w_1$ at the first time point (corresponding to the addition) and $w_2$ at the second time point. 
When the pixel activation ratio is $p$, the probability of the first MAC operation occurring as scheduled is $p$. When it is pruned, the following MAC operations are shifted earlier on the timeline. 
Subsequently, for the $j$-th MAC operation ($j>1$), it executes with a probability $p$ and its execution position spread out according to which of the $2^{j-1}$ possible sequences for the previous $j-1$ operations is executed.
The highest probability for one sequence containing the $j$-th MAC operation executing at a particular time point is $p \cdot \max(p,1-p)^{j-1}$, as derived later in Section~\ref{sec:theory}.
\fei{I find this position analysis not correct. the $j^{th}$ operation only has j positions, right?} 
\ruyi{First there is an assumption that jth operation is execute (the first p) in the formula, then, for all operations before j, (there are j-1) operations, we can execute or not. So the number is $2^{j-1}$. Please note that we have another assumption, if the activation sequence is different, the leakage point will be different.}

Mangard et al.~\cite{mangard2008power} demonstrated that, with uncertain leaky time points as a countermeasure, the required \update{number of traces for successful secret retrieval increases by $R$ times over the original traces required}, with $R = \hat{p}^{-2}$, where $\hat{p}$ is the probability of a specific operation occurring at a predetermined time.
Here we follow this notation: $R$ is defined as the ratio between the number of traces required to retrieve the secret in a protected system versus unprotected ones using DEMA.
Thus for attacking the $j$-th weight at the best time point, $R = p^{-2} \cdot \max(p,1-p)^{2(j-1)}$. For DNNs with a lot of parameters, the challenge of conducting side-channel attacks on later parameters is amplified due to the uncertain time of operation executions, and the attack complexity increases exponentially with $j$.
\ding{We may also want to start the $p$ in Table 1 from $p=0.5$ only. Or change the $p<0.5$ cases to the new formula }
 \ruyi{I update the table with $p<0.5$ using new equation}

\begin{table}[h]
  \centering
  \small
    \caption{Leaky time point $j^*$ when $R\geq1000$ for ratio $p$}
  \begin{tabular}{c|c|c|c|c|c|c|c|c|c}
\hline
    $p$   & 0.1 & 0.2 & 0.3 & 0.4 & 0.5 & 0.6 & 0.7 & 0.8 & 0.9 \\
    \hline
    $j^*$ & 12   & 10   & 8   & 6   & 5   & 7   & 10  & 16  & 33  \\
    \hline
  \end{tabular}

  \label{tab: Nr vs p}
\end{table}

We define $j^*$ as the earliest $j$th time point when $R\geq 1000$. This is based on the premise that attacks on the original design, vulnerable with just $1000$ traces, become impractical on the system protected by \method when $R \geq 1000$, i.e., requiring over a million traces. 
As trace collection is time-consuming in EM and power side-channel analyses, a significant increase in trace count deters such attacks and one million traces is a common threshold for side-channel power/EM attacks. 
A smaller $j^*$ indicates better protection on the DNN.
Table~\ref{tab: Nr vs p} shows how the activation ratio $p$ affects the value of $j^*$: 
when $p=0.5$, we have $j^*=5$, which implys that only the first $5$ weights remain vulnerable. 
Considering the large number of weights in DNN layers (e.g., 864 weights in MobileNetV2's first layer), exposure of a small subset does not significantly compromise the network confidentiality.
Therefore, \texttt{RPAM} is deemed an effective mitigation for SCA on embedded DNNs.

\subsection{Importance-aware Pixel Activation Map} \label{sec:ipam}

\begin{figure}[t]
\centering
\includegraphics[width=\linewidth]{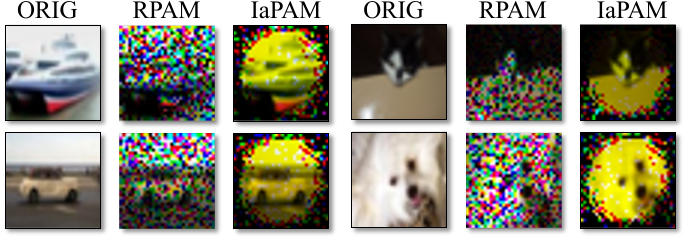}
\caption{A CIFAR-10 image with activation ratio $0.5$}
\label{fig: two masks}
\end{figure}

Random activation of input pixels works well for simple datasets such as MNIST. However, when input is complex, i.e., CIFAR-10, the information loss significantly impairs the performance of pre-trained models.
Fig.~\ref{fig: two masks} illustrates this phenomenon (the second and fifth columns), where randomly activated pixels (\texttt{RPAM}) fail to preserve crucial details of the input (under activiation rate of 50\%).
To address such degradation, we leverage the intrinsic nature of DNN inference--input features vary in their contributions to the model performance: more informative features such as the object's shape and colors often directly affect the prediction outputs; while features with less information, such as the background, might have much less influence on the results~\cite{selvaraju2017grad}.
In light of this, we introduce the Importance-aware Pixel Activation Map (\texttt{IaPAM}), which enhances the robustness of the protected DNN by preventing the most critical pixels in the input based on their importance to the model's performance from being dropped, i.e., never deactivated.
Fig.~\ref{fig: two masks} (the third and sixth columns) shows the result of applying \texttt{IaPAM}, where the principal patterns of the input image are well-preserved (near the center), and other less important pixels are randomly dropped. 
We expect that its classification accuracy will exceed that of \texttt{RPAM}.

\begin{figure*}[t]
\centering
\includegraphics[width=\linewidth]{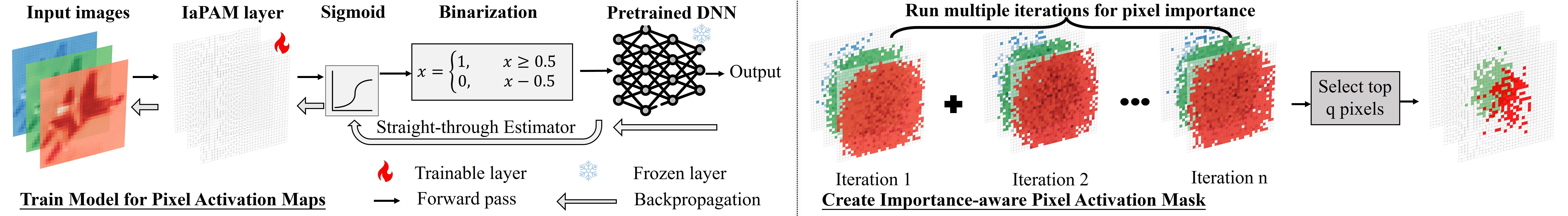}
\caption{\texttt{IaPAM} recognizes the variance of input pixel importance and finds the most critical pixels to generate an activation map with STE and one-shot pruning.}
\label{fig: importance-aware Activation Mask}
\end{figure*}

The \texttt{IaPAM} is a trainable activation map applied to the inputs of DNN. It is integrated as an additional layer preceding the pre-trained model, denoted $\mathcal{M}$. 
Fig.~\ref{fig: importance-aware Activation Mask} illustrates the training process.  
Without loss of generality, assuming the input is two-dimensional ($[I, J]$), the coordinate of a pixel in the input is $(i, j)$, and the weights associated with each pixel are represented by $m_{ij}$.
These weights $m_{ij}$ quantify the relative importance of each pixel, and are modulated by a sigmoid function to constrain their output range.
Similar to \texttt{RPAM}, we define $p$ as the proportion of pixels activated and $q$ as the maximum proportion of critical pixels, thereby the uncritical weights will be randomly activated with a probability at $\frac{p-q}{1-q}$.
Given the non-differentiability of the binary map during back-propagation, we employ the straight-through estimator (STE) technique, as suggested by~\cite{ courbariaux2015binaryconnect}, to facilitate the optimization of the activation map's weights.
\fei{zero means critical or one means critical?} \ruyi{one mean critical}

The loss function for IaPAM training utilizes an L1 regularization:

\begin{equation}
L_{\texttt{IaPAM}} = L_{CE} + \alpha L_1(\frac{\text{nnz}(\mathcal{M})}{|\mathcal{M}|}-q)
\label{eq: loss for LaPAM}
\end{equation}

Here, $L_{CE}$ denotes the cross-entropy loss for classification, the L1 regularization term controls the difference between the activated pixels in $\mathcal{M}$ and the desired ratio $q$, where the operator $\text{nnz}(\cdot)$ counts the non-zero values and $|\cdot|$ compute the cardinality. 
We utilize the coefficient $\alpha$ to strike a balance between the DNN performance and the input sparsity, which will be further studied in Section~\ref{sec: ablations}.
The full algorithm is shown in Algorithm~\ref{alg: iapam}. 
We optimize the map layer $\mathcal{M}$ for multiple iterations and compute an importance score for each pixel $\mathcal{S}$ by summing up the associated weights importance for every iteration.
The pixels corresponding to the top $q$ importance scores are selected as the critical ones.
Although a subset of pixels is always activated, the remaining pixels are randomly activated at a scaled ratio, further complicating operation localization by the adversary. For example, take a scenario where $p=0.8$ and $q=0.5$; here, non-critical pixels are randomly activated at a ratio of $\frac{p-q}{1-q}=0.6$.
Referring to Table~\ref{tab: Nr vs p}, when $p=0.8$, if the initial 16 pixels are non-critical, $R$ will exceed 1000 at the 17th weight.\fei{when checking Table 1, should you use p = 0.8 or 0.6? I think you should still use p=0.8?}
\ruyi{This will be more complex to do mitigation analysis IaPAM. If there is critical pixel in the first 7 pixels (consider the fact that most of critical pixels are centralized, the mitigation strength will equal to p=0.6. }
As seen in Figure~\ref{fig: masks-for-cifar} (in Section~\ref{sec: exp: IaPAM}), the upper left corner where the computation starts is always non-critical for images in CIFAR-10, thus the weight confidentiality is still well-preserved.

\begin{algorithm}[t]
\caption{Importance-aware Pixel Activation Map}\label{alg: iapam}
\begin{algorithmic}[1] %
\REQUIRE Pre-trained model $f$, Iterations number $Iter$, Critical weights ratio $q$, Training Dataset $(x_\text{train}, y_\text{train})$, Number of Epoch $E$, Map weights $\{m_{ij}\}$
\ENSURE Importance-aware Pixel Activation Map $\mathcal{M}$.

\STATE Initialize $m_{ij}\leftarrow\mathbf{0}$
\FOR{ $\text{iter}=1, 2, \dots, Iter$}
    \FOR{$epoch=1, 2, \dots, E$}
        \STATE Activate the map $\mathcal{M}=\{\text{sigmoid}({m_{ij}})\}$
        \STATE Do inference with map $y_{\text{pred}}=f(\mathcal{M}(x_\text{train}))$
        \STATE Compute the Cross-entropy Loss $L_{\text{CE}}(f(x_{\text{train}}), y_{\text{train}})$
        \STATE $\{m_{ij}\}\leftarrow$ optimize the map weights with $L_\text{IaPAM}$~\eqref{eq: loss for LaPAM}
    \ENDFOR
    \STATE Update the important score of each pixel: $\mathcal{S} = \mathcal{S} + \{\text{sigmoid}(m_{ij})\}$ 
\ENDFOR
\STATE $\mathcal{M}_{ij}\leftarrow \begin{cases} 1 & \text{if } \mathcal{S}_{ij} \text{ is the top } q \text{ percentage values in } \mathcal{S}\\ 0 & \text{otherwise} \end{cases}$
\RETURN $\mathcal{M}$
\end{algorithmic}
\end{algorithm}

\section{Mitigation Strength Analysis}\label{sec:theory}

\ding{Above I explained how the formulas were derived. I suggested the following revision, which may not exactly correspond to what Cheng reported, but I think we should keep it simple, instead of jumping around quantities $\frac{\mathrm{Cov}(l_m,v_m)^2 \sigma'^4}{\sigma^4\mathrm{Cov}(l_m',v_m)^2 }$, $\frac{\mathrm{Cov}(l_m,v_m)^2 \sigma'^2}{\sigma^2\mathrm{Cov}(l_m',v_m)^2 }$ or  $\frac{\mathrm{Cov}(l_m,v_m)^2 }{\mathrm{Cov}(l_m',v_m)^2}$. Basically Cheng made the assumption that the noise $\sigma^2 \approx \sigma'^2$, thus they all are about the same}

To demonstrate the effectiveness of our proposed \method in defending against DEMAs, we analyze its theoretical mitigation strength, specifically, apply \texttt{RPAM} to DNN operations, and empirically verify it with results presented in Section~\ref{exp: empirical}.

We continue using the metric $R = \frac{N}{N_0}$
which represents a ratio between the current number of traces required for successful secret retrieval $N$ with \method and the original number of trace requirements $N_0$.
A higher $R$ indicates more traces are required to recover the same secret information than before, and $R=1,000$ is treated as the threshold of a successful defense.

\subsection{Theoretical Formula Derivation of $R$} \label{sec: theoretical R}
Assume we aim to attack the operation at $j$-th MAC for the unprotected DNN model. 
First, this operation will be preserved with a probability of $p$, i.e., among $N$ traces used to do SCA on this operation, $p \cdot N$ traces will contain the valid information. 
However, because the random MAC drop also occurs in all early $j-1$ MAC operations, the leakage is distributed among $2^{j-1}$ possible sequences of operations kept by \method. %
The varying sequences also lead to the leakage related to the $j$-th original MAC operation occurring at varying time points.
This is indeed the leakage hiding effect introduced by the \method to mitigate side-channel attacks. 

In DEMA, we assume the attacker will examine the collected EM traces and find the largest leakage point.
Our theoretical leakage signal degradation analysis will be on the largest leakage time point.
Any specific sequence of $k-1$ operations before the $j^{\text{th}}$ MAC will occur $p^k(1-p)^{j-k}$ proportion of times.
Thus we have the highest proportion of original $j$-th MAC leakage preserved through a sequence of (containing this MAC) operation as
$$\max_{1 \le k \le j}p^k(1-p)^{j-k} = \max_{1 \le k \le j} p^j (\frac{1-p}{p})^{j-k} $$

When $p \ge 0.5$, we have $\frac{1-p}{p} \le 1$, thus the maximum of the above expression is $p^j$ achieved at $k=j$ (i.e., all $j$ operations are kept); when $p<0.5$, $\frac{1-p}{p} > 1$, the maximum value is $p(1-p)^{j-1}$ achieved at $k=1$ (i.e., all first $j-1$ operations are dropped, and the $j$-th operation is executed). 
Combining those two cases, for all $p$ values, the maximum amount of leakage among all times points is $p \cdot \max(p,1-p)^{j-1}$ proportion of original leakage. 

To derive the ratio of the number of traces required $R$ from the max leakage signal level, we utilize the signal-to-noise ratio ($SNR$), defined as the ratio between the variance of the leakage signal and the variance of noise at the leakage time point. 
When leakage signal becomes smaller by a ratio of $p \cdot \max(p,1-p)^{j-1} $, its variance is reduced by a ratio of $p^2 \cdot \max(p,1-p)^{2(j-1)}$. 
Since the noise level is not affected by \method, the SNR is also reduced by a ratio of $p^2 \cdot \max(p,1-p)^{2(j-1)}$. 
$SNR$ is related to the number of trace requirement~\cite{mangard2008power}, i.e., $N \propto \frac{1}{\mathrm{SNR}}$, indicating that change in $N$ is inversely proportional to $SNR$ change.
Hence the theoretical R is :
\begin{equation}
    \label{eq: R}
    R=p^{-2}\max(p,1-p)^{-2(j-1)}
\end{equation}

\begin{figure}[t]
    \centering
    \includegraphics[width=0.8\linewidth]{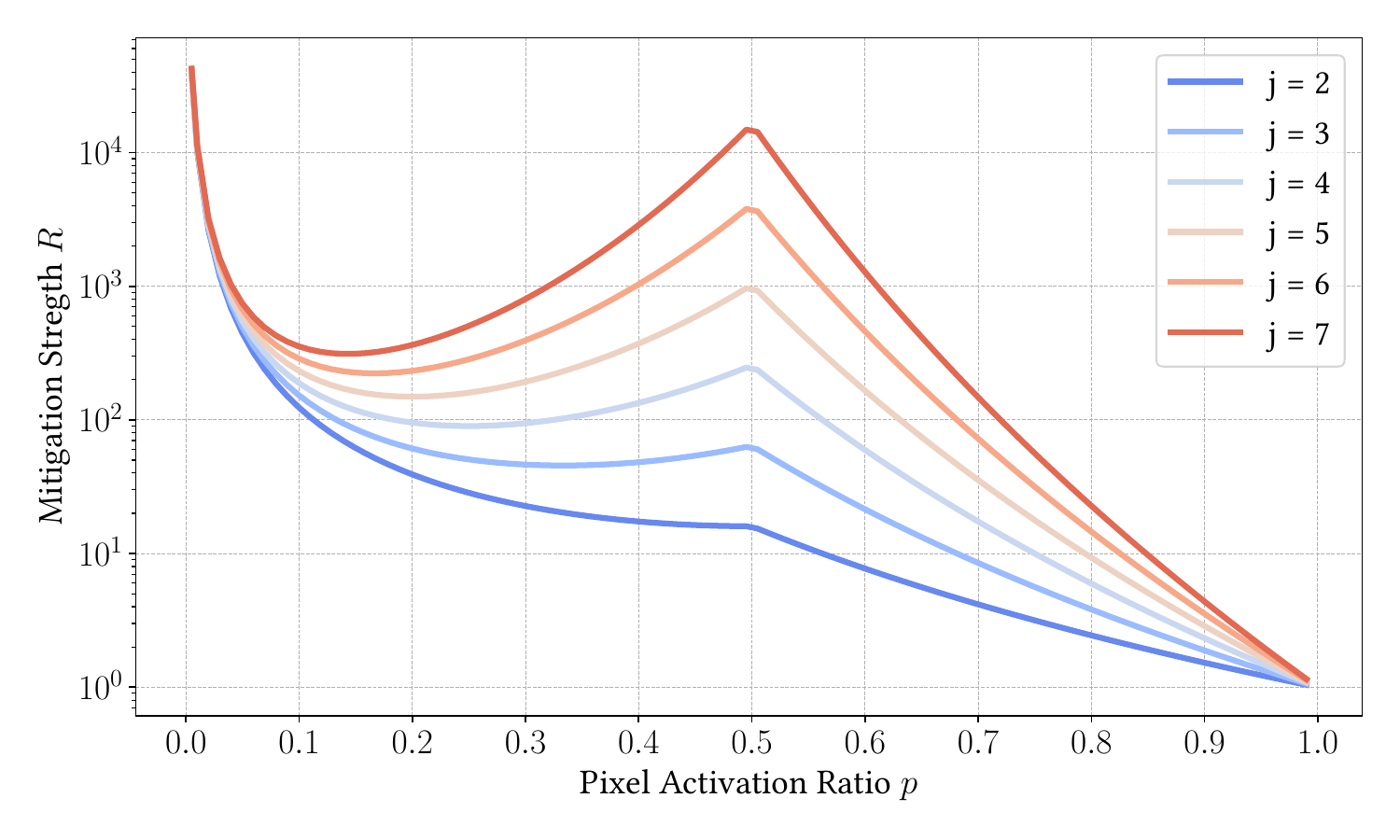}
    \caption{Visualize Function $R$ under different $j$}
    \label{fig: visualize R}
\end{figure}

According to Formula~\eqref{eq: R}, when we focus on protecting the $j$-th operation weight for $j\le 2$, decreasing $p$ always increases $R$, leading to enhanced protection. However, for $j \ge 3 $, as shown in Fig.~\ref{fig: visualize R}, $R$ increases when $p$ decreases from $1$ to $0.5$, then $R$ decreases when $p$ decreases from $0.5$ to $1/j$, before $R$ increases again when $p$ further decreases from $1/j$. For the later weights with a large $j$ value, unless we use a very small $p<1/j$, the optimal protection is achieved at $p=0.5$. Since too small $p$ will degrade the resulting model performance too much (will be presented in Table~\ref{tab: IaPAM performance}) and thereby should be excluded, $p=0.5$ would be an optimal choice for protection in most practical applications. Thus, when considering the trade-off between protection and the model performance, we can focus on choosing among values of $p \ge 0.5$.

\subsection{Formula of $R$ for Empirical Measurements}
To verify Formula~\eqref{eq: R} experimentally, we estimate the $SNR$ with and without \method. 
 We first identify the strongest leakage point for the $j$-th operation by profiling, using the point with the highest Pearson Correlation Coefficient with the correct weight value.
Then, we estimate the ratio of $SNR$ change by leveraging a statistical model of EM side-channel leakage~\cite{fei2015statistics}:
\begin{equation}
    \label{eq: EM model}
    l_m = \epsilon v_m + r_m, \quad m=1,...,n.
\end{equation}
Here, $l_m$ denotes the EM measurement at a specific time point, $\epsilon$ is a constant representing the unit EM emanation, $v_m$ is the selection function (known as the leakage model, where we use the HW of the intermediate accumulation), and $r_m$ is the random noise influenced by various factors such as measurement, electrical, and switching noise, following a Gaussian distribution $N(c,\sigma^2)$. Then, $SNR \propto \epsilon^2/\sigma^2$. Hence, we can estimate $R$ using the equation:
\begin{equation}
    \hat R  = \frac{\mathrm{\epsilon^2 (\sigma')^2}}{(\epsilon')^2 \sigma^2},\label{equ:R.emp}
\end{equation}
where $\epsilon'$ and $\sigma'$ represent the post-mitigation values when fitting \eqref{eq: EM model} on measurements with \method activated. 

In Fig.~\ref{fig: Nr}, we plot the theoretical $R$ values Formula~\eqref{eq: R} versus the empirical estimates~(\ref{equ:R.emp}). They exhibit a good match, affirming the validity of the theoretical model.
These $R$ values demonstrate the effectiveness of the \method in mitigating side-channel analysis for each weight, with the effectiveness increasing exponentially for subsequent operations (linear increase in the logarithm scale plot).

\begin{figure}[t]
    \centering
    \includegraphics[width=\linewidth]{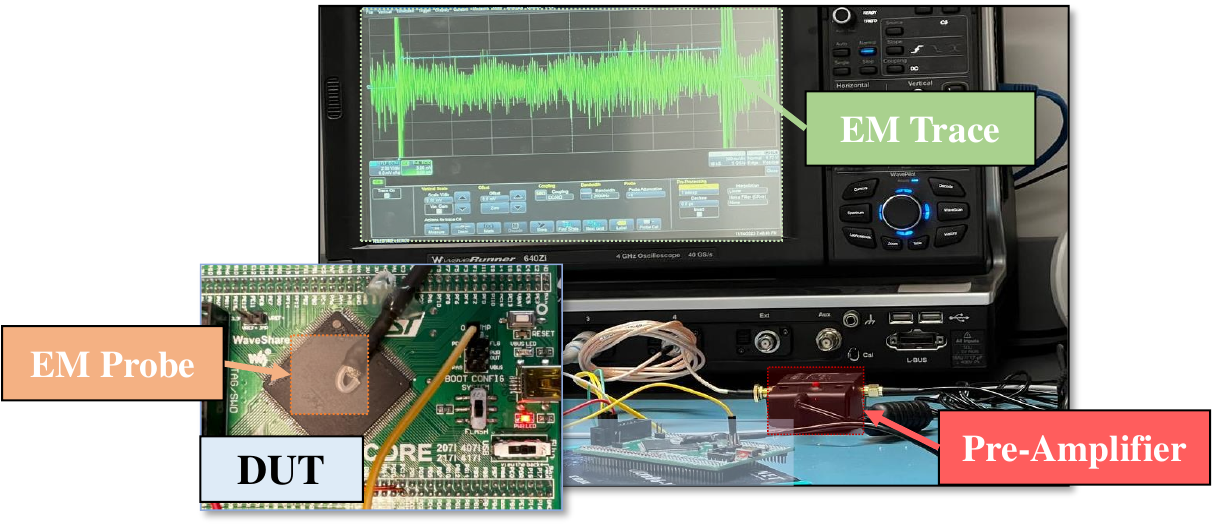}
    \caption{Setup of EM leakage measurements.}
    \label{fig: experiment setup}
\end{figure}

\section{Evaluation}

\subsection{Experiment Setup} \label{sec: exp: setup}
\subsubsection{Measurement Setup}
Fig.~\ref{fig: experiment setup} depicts the trace collection setup.
The DUT has an ARM Cortex-M4 \verb|STM32F417IG| MCU\footnote{\url{https://www.st.com/en/microcontrollers-microprocessors/stm32f417ig.html}} featuring DSP and FPU capabilities, whose operating frequency is 168~\unit{MHz}.
An EM probe coupled with a pre-amplifier from Aaronia AG\footnote{\url{https://aaronia.com/en/probe-set-pbs-2-inkl-vorverstaerker}} is utilized to capture EM signals and convert them into voltage readings.
This probe is strategically positioned above the emission zones of the DUT and is linked to a Lecroy oscilloscope\footnote{\url{https://teledynelecroy.com/oscilloscope/}} via the pre-amplifier.
Both the DUT and the oscilloscope are interfaced with a host workstation, facilitating command execution and synchronization.
To acquire an EM trace, the host initiates the process by instructing the DUT to commence DNN inference.
Concurrently, the oscilloscope records the EM signals at a sampling rate at 5~\unit{GS/s}, which are then streamed to the workstation.
\subsubsection{Model Training Setup}
We evaluate \method's performance using MobileNet-V2~\cite{sandler2018mobilenetv2}, a widely adopted DNN for on-device applications on MNIST and CIFAR-10.
We train the MNIST model from scratch, allocating $80\%$ of the data for training and $20\%$ for testing, achieving a testing accuracy of $99.35\%$.
For CIFAR-10, we utilize a pretrained MobileNet-v2\footnote{\url{https://github.com/huyvnphan/PyTorch_CIFAR10}} with an accuracy of $93.15\%$.

\subsection{Selection of Leakage Points}
\begin{figure}[t]
    \centering
    \includegraphics[width=\linewidth]{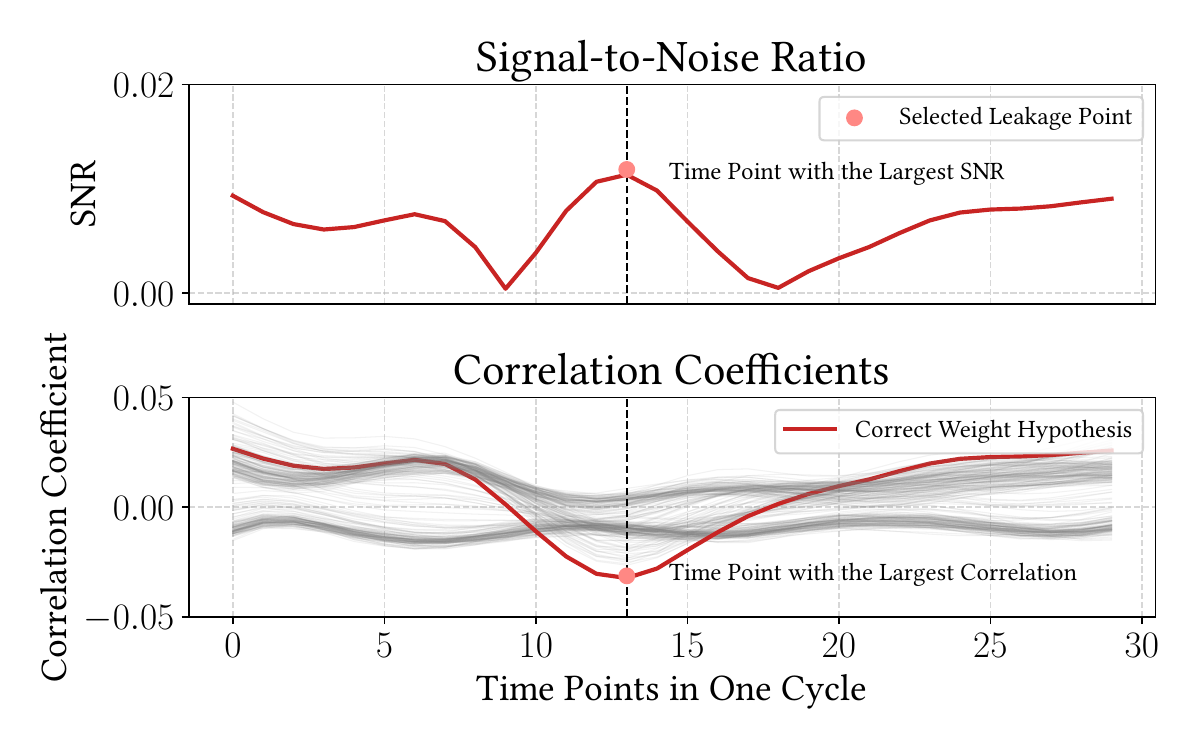}
    \caption{Select Leakage Point(s) via SNR and Correlation}
    \label{fig: leakage points}
\end{figure}

To assess the mitigation strength fairly, we select time points exhibiting the strongest leakage for each MAC operation. 
The underlying rationale is straightforward: if an attacker cannot exploit the time point with the highest SNR, they cannot possibly exploit less informative ones. 
Specifically, the execution of assembly code by MCUs is deterministic, lacking advanced features such as branch prediction, out-of-order execution, or prefetching. 
Consequently, each MAC operation produces EM traces of uniform length, facilitating segmentation. We identify the clock cycle where leakage occurs and select this cycle for analysis, empirically confirming it exhibits the strongest leakage. 
Knowing the correct secret weight for each MAC operation enables us to compute the SNR for each time point in this segment using Formula~\eqref{eq: EM model}, as depicted in Fig.~\ref{fig: leakage points}. 
We target the time point with the highest absolute SNR value. 
For comparative analysis, we also display the DEMA results, where the Pearson Correlation coefficient for the correct guess is marked in {\color{red!60} red}, while other guesses are {\color{gray!50} greyed} out. 
The analysis shows that the time point with the strongest SNR corresponds to the DEMA results, which could be used by attackers to precisely extract secrets.

\subsection{Empirical Mitigation Strength on \texttt{RPAM}} \label{exp: empirical}
\begin{figure}[t]
    \centering
    \includegraphics[width=\linewidth]{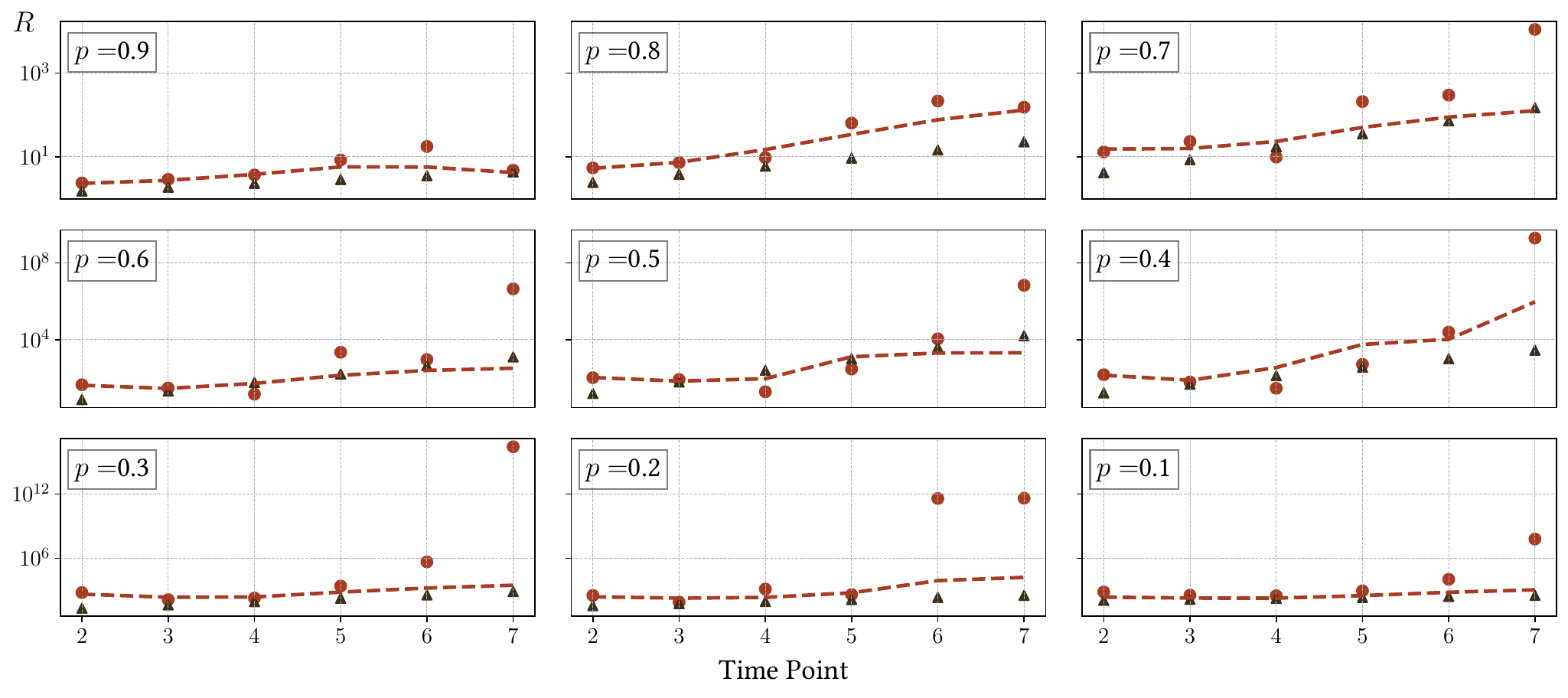}
    \caption{\textbf{Empirical measurement of $R$.} The symbol $\textcolor{mes_color}{\bullet}$ represents the measured values, while $\textcolor{theo_color}{\blacktriangle}$ represents the theoretical values.
    }
    \label{fig: Nr}
\end{figure}

The results of the empirical measurement are illustrated in Fig.~\ref{fig: Nr}. The x-axis represents the time point, corresponding to individual MAC operations. We initiate the count from two to mitigate the impact of HW aliasing discussed in Section~\ref{sec: observations}.
The y-axis displays $R$ in log scale for enhanced clarity. 
The symbol $\textcolor{theo_color}{\blacktriangle}$ denotes the theoretical protection strength (Equation \eqref{eq: R}), while $\textcolor{mes_color}{\bullet}$
signifies the average of 20 separate measurements, with measurement is calculated on 100,000 EM traces (Equation \eqref{equ:R.emp}). 
The experiment demonstrates that our empirical measurements fit well with the theoretical mitigation strength for all $p$ values analyzed. 
In particular, with a smaller $p$, both theoretical and empirical mitigation strength increase, and for the pixels executed later (those with a large $j$ index), the protection will be better, which perfectly matches our analysis in Section~\ref{sec:theory}. However, an excessively small $p$ might cause significant model accuracy degradation, which will be further discussed in Table~\ref{tab: IaPAM performance}.

\ding{what are the theoretical value $\textcolor{theo_color}{\blacktriangle}$ and $\textcolor{mes_color}{\bullet}$? I thought they were values from \eqref{eq: R} and \eqref{equ:R.emp} respectively. Is that true? You used 20 datasets repeatedly fitting \eqref{equ:R.emp}?}

\subsection{Evaluation on \texttt{IaPAM}} \label{sec: exp: IaPAM}

\begin{figure}[h]
    \centering
    \includegraphics[width=\linewidth]{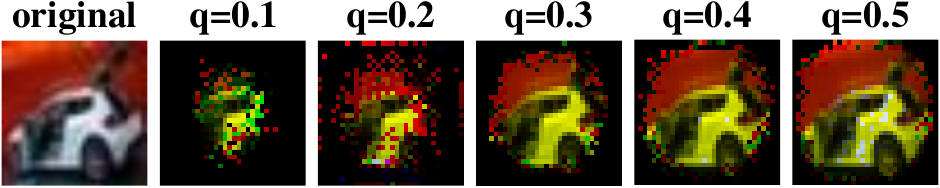}
    \caption{Obtained critical pixels at different $q$ values.}
    \label{fig: masks-for-cifar}
\end{figure}

The proposed \texttt{IaPAM} identifies and maintains the activation of critical pixels in input images throughout the inference process.
Fig.~\ref{fig: masks-for-cifar} illustrates the masked CIFAR-10 images at various importance activation ratios, referred to as $q$.
Notably, \texttt{IaPAM} operates on the individual channels of the 3-channel RGB inputs.
It consistently highlights that vital pixels tend to cluster around the central area of the inputs, which frequently corresponds to the main subject of the classification task.
Such a concentration of importance scores of \texttt{IaPAM} in the center significantly enhances the model's defense performance against side-channel attacks targeting DNN parameter extraction, which usually starts at randomly activated pixels on the peripherals, making it challenging for an adversary to sequentially extract weights effectively.
Moreover, our analysis indicates a tendency for importance scores to concentrate more on the Red and Green channels of CIFAR-10 inputs.
This is presumably because elements in the Blue channel, such as the sea and sky, often represent background areas that are less crucial for classification purposes.

In Table~\ref{tab: IaPAM performance}, we demonstrate the superior effectiveness of \texttt{IaPAM} (the third to seventh column with different $q$), compared to \texttt{RPAM} (the second column).
To quantify the impact of additional pixel activation maps on model accuracy, we introduce the metric of relative accuracy drop, defined as $\frac{\text{Acc}_{\text{org}} - \text{Acc}_{\text{pam}}}{\text{Acc}_{\text{org}}}$, where $\text{Acc}_\text{org}$ is the accuracy of the plain model without \method, and $\text{Acc}_\text{pam}$ is that of the protected secure model.
Notably, \texttt{IaPAM} significantly mitigates the performance degradation typically associated with the loss of information when inputs are randomly discarded.
For instance, when the activation ratio is $70\%$, the degradation of \texttt{IaPAM} with $q$=40\% is $2$ times smaller than that of \texttt{RPAM}.

\begin{table}[t]
    \centering
    \footnotesize
    \caption{Relative accuracy drop by applying different activation strategies during DNN inference on CIFAR-10. $p$ denotes the activation ratio and $q$ denotes the ratio of critical pixels.}
    \resizebox{\linewidth}{!}{
        \begin{tabular}{c|c|c|c|c|c|c}
            \toprule[0.3mm]
            $p$ & \texttt{RPAM} & $q$=0.1 & $q$=0.2 & $q$=0.3 & $q$=0.4 & $q$=0.5 \\
            \hline
            0.2 & 15.08\%$\downarrow$      & 7.84\%$\downarrow$ & -       & -       & -       & -       \\
            \hline
            0.3 & 10.20\%$\downarrow$      & 6.72\%$\downarrow$ & 5.24\%$\downarrow$  & -       & -       & -       \\
            \hline
            0.4 & 7.64\%$\downarrow$       & 6.54\%$\downarrow$  & 5.01\%$\downarrow$ & 4.75\%$\downarrow$  & -       & -       \\
            \hline
            0.5 & 5.35\%$\downarrow$       & 5.03\%$\downarrow$  & 4.07\%$\downarrow$  & 4.05\%$\downarrow$  & 2.23\%$\downarrow$  & -       \\
            \hline
            0.6 & 4.30\%$\downarrow$       & 4.17\%$\downarrow$  & 3.23\%$\downarrow$  & 3.17\%$\downarrow$  & 1.81\%$\downarrow$  & 1.18\%$\downarrow$  \\
            \hline
            0.7 & 3.48\%$\downarrow$       & 2.66\%$\downarrow$  & 2.64\%$\downarrow$  & 2.52\%$\downarrow$  & 1.72\%$\downarrow$  & 0.91\%$\downarrow$  \\
            \hline
            0.8 & 2.13\%$\downarrow$      & 2.04\%$\downarrow$  & 1.96\%$\downarrow$  & 1.76\%$\downarrow$  & 1.05\%$\downarrow$  & 0.68\%$\downarrow$  \\
            \hline
        \end{tabular}}

    \label{tab: IaPAM performance}
\end{table}

\subsection{Overhead and Comparison}
To assess the execution overhead of \method on MCUs, it is crucial to examine its implementation details, especially the MAC operations within a loop structure. \update{The integration of our defensive strategy necessitates the inclusion of a conditional check before each MAC operation to determine whether it should be executed or skipped, based on a random number. This random number is held in a 32-bit unsigned integer and is accessed using a one-bit map that shifts left with each iteration.}\fei{is this implementation only for RPAM? how to implement Ruyi's IaRPAM.}
The Cortex-M4 programmer's manual~\cite{cortex-M4} shows that the branching instruction of this check takes one cycle if not executed, and $1+D$ cycles otherwise, with $D$ being the pipeline length (1-3 stages). Random number generation involves a left shift and `and' operation, each taking one cycle. Index updates require two cycles, and two loads need three cycles. Assuming a DSP extension, a MAC operation takes one cycle. For a first layer with $M$ MAC operations, the total cycles under defense of \texttt{RPAM} are $Mp(1+D+2+6)+M(1-p)=((8+D)p+1)M$, versus $6M$ without defense. For the defended system to be faster, the pixel activation ratio $p$ must satisfy $p < \frac{5}{8+D}$. Even at $D = 3$, the system is faster if $p < 0.45$. This aligns with measurement results on our DUT with $D=3$.
Implementing \texttt{IaPAM}, akin to \texttt{RPAM}, involves an extra conditional instruction to assess pixel importance. This process includes a 5-cycle lookup operation to identify critical pixels using a binary table.  An additional conditional branch is used to decide on random pixel activation. The total number of cycles for \texttt{IaPAM} for the first layer is $[(8+D)p+13-8q]M$.
\ding{This is overhead: after subtracting 6M?}\ruyi{Yes, revised here.}

According to results shown in Fig.~\ref{fig: visualize R}, an ideal random pixel activation ratio is $0.5$. 
Therefore, we take the performance of the DNN overhead and accuracy degradation on CIFAR-10 for \texttt{IaPAM} when $p=0.7$ and $q=0.4$, i.e., $\frac{p-q}{1-q}=0.5$ and the \texttt{RPAM} with the same activation ratio $p=0.7$, as shown in Table~\ref{tab: different models}.
We also do the same evaluation on multiple models including MobileNet-v2~\cite{sandler2018mobilenetv2}, DenseNet121~\cite{huang2017densely}, and GoogleNet~\cite{szegedy2015going}, both methods show effectiveness in balancing accuracy drop and overhead.
Compared with the reported $18\%$ overhead for state-of-the-art (SOTA) shuffling defenses~\cite{brosch2022counteract} and $127\%$ for the masking on DNN~\cite{dubey2020maskednet}, both \texttt{RPAM} and \texttt{IaPAM} outperform in the computation overhead with acceptable accuracy degradation.

\begin{table}[t]
    \centering
    \footnotesize
        \caption{Accuracy Degradation and Defense Overhead}
    \resizebox{\linewidth}{!}{
        \begin{tabular}{l|c|c|c|c|c}
            \toprule[0.3mm]
            \multirow{2}{*}{Structure} & Original & \multicolumn{2}{c|}{\texttt{RPAM} (p=0.7)} & \multicolumn{2}{c}{\texttt{IaPAM} (q=0.4, p=0.7)} \\\cline{3-6}
              & Accuracy & Acc. Drop                & Overhead         & Acc. Drop               & Overhead          \\\hline
            \hline
            MobileNetV2 & 93.15\%  & 3.48\%$\downarrow$ & 0.27\%$\uparrow$ & 1.72\%$\downarrow$ & 0.61\%$\uparrow$ \\
            DenseNet121 & 94.06\%  & 5.71\%$\downarrow$ & 0.26\%$\uparrow$ & 2.74\%$\downarrow$ & 0.58\%$\uparrow$  \\
            GoogleNet   & 92.84\%  & 3.78\%$\downarrow$ & 1.01\%$\uparrow$ & 2.87\%$\downarrow$ & 2.23\%$\uparrow$  \\
            \hline
        \end{tabular}}

    \label{tab: different models}
\end{table}

\subsection{Ablation Study} \label{sec: ablations}
\begin{figure}[t]
    \centering
    \begin{subfigure}[b]{0.49\linewidth}
    \centering
        \includegraphics[width=0.99\linewidth]{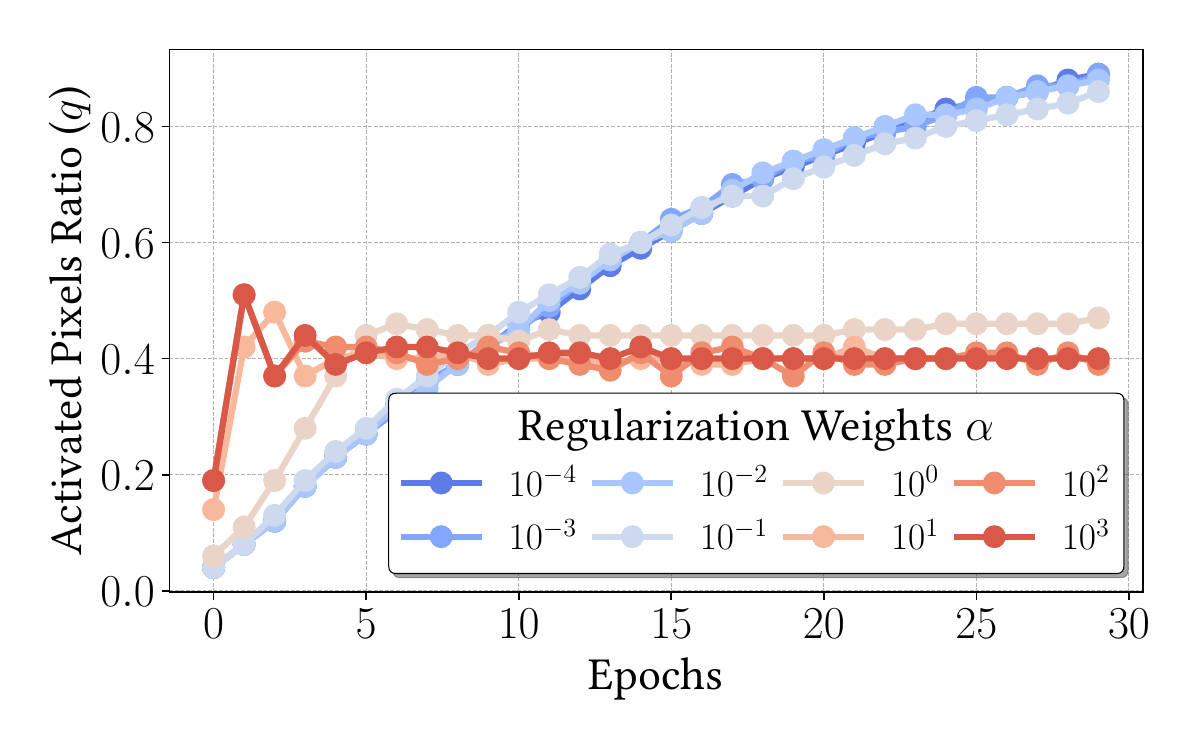}
        \caption{Epochs vs Activated Ratio}
        \label{fig:alpha-sub1}
    \end{subfigure}
    \hfill 
    \begin{subfigure}[b]{0.49\linewidth}
        \centering
        \includegraphics[width=0.99\linewidth]{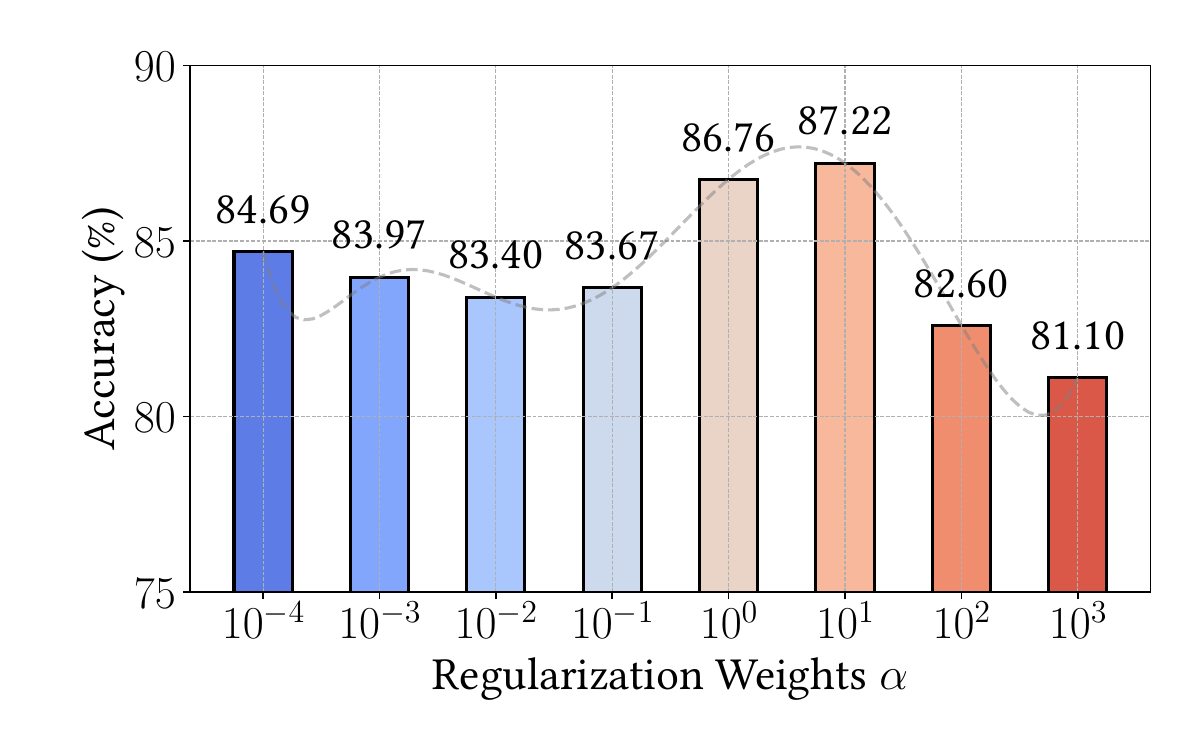}
        \caption{IaPAM Performance vs $\alpha$}
        \label{fig:alpha-sub2}
    \end{subfigure}
    \caption{Ablation Study on the Regularization Weights $\alpha$}
    \label{fig: ablation study on alpha}
\end{figure}

In the design of IaPAM, the regularization term $\alpha$ in $L_{\text{IaPAM}}$ is utilized to balance the map activation ratio for each iteration (controlled by the $L_1$ norm) and the deactivated pixel's importance (controlled by $L_{CE}$).
To further understand the effectiveness of regularization weights $\alpha$, we evaluate the IaPAM performance at different $\alpha$ values. 
In our experiment, we select $p=0.7$ and $q=0.4$. In Fig.~\ref{fig:alpha-sub1}, we visualize the activation ratio change with the training epochs with different $\alpha$, and in Fig.~\ref{fig:alpha-sub2}, we show the IaPAM performance with various $\alpha$ when fine-tuning the pre-trained model with $5\%$ training samples.
Note that we use a small number of training samples to fine-tune the pre-trained model so as to demonstrate the impact of different $\alpha$ more clearly.

\noindent\textbf{Activated Pixel Ratio versus $\alpha$:}
The choice of $\alpha$ will directly affect the final number of activated pixels during training of the mapping layer $\mathcal{M}$.
In particular, we observe that the activation ratio will increase from $0$ to the final activated ratio as the map is initialized as all-zero.
A small choice of $\alpha$ will cause the cross-entropy $L_{CE}$ to dominate the total loss, causing the activated ratio to keep increasing and exceed the maximum threshold of $q$ (as shown in the {\color{bleudefrance!60} blue} curves in Figure~\ref{fig:alpha-sub1}).
On the other hand, a higher $\alpha$ (the {\color{red!60} red} curves) will ensure the final activated pixel ratio is close to $q$.

\noindent\textbf{IaPAM performance versus $\alpha$:}
From Fig.~\ref{fig:alpha-sub2}, we observe that neither a too-small $\alpha$ nor too-large one will result in a good performance for IaPAM.
Specifically, according to Algorithm~\ref{alg: iapam}, as the final activated pixels ratio is higher than $q$, we will compute the importance scores based on the sum from multiple iterations.
In this case, critical weights might be removed, causing the final accuracy to be lower due to loss of information.
On the other hand, when $\alpha$ is large, the optimization process focuses more on the sparsity rather than the importance of pixels to minimize $L_{CE}$, thus causing poor performance after fine-tuning.
In conclusion, we carefully select $\alpha$ to ensure a balance between $L_{CE}$ and the $L_1$ regularization term.
A selection of $\alpha=1$ is used in the paper.

\section{Adaptive Attack against \method} \label{sec: adaptive}

\method mitigates side-channel-based weight-stealing attacks by redistributing the leakage signals of DNN weights across different time points. We assume the attacker targets the Hamming weight model of cumulative results, requiring weights to be attacked sequentially. So far the attack model is to exploit the leakage at the
$j$-th time point corresponding to the $j$-th weight. %
\method mitigates the attack effectively, as the probability of obtaining traces with meaningful leakage signals from the targeted attack model diminishes exponentially as $j$ increases.

In this section, we consider a stronger attacker who is aware of \method and can utilize all leakage information at a certain time point across various sequences containing the target operation. While combining leakage signals is inherently challenging and worth further exploration, in this work we focus on analyzing the upper bound of the leakage signal for any potential adaptive attack, providing guidance for follow-on pragmatic attacks.

\subsection{Threat Model for Potential Adaptive Attack}
Similar to previous setting, we assume the attacker knows the DUT implementation details including the sequence of MAC operations. 
The attacker can control the inputs to the edge DNN and capture the EM emissions from the DUT.
Previously our analysis focused on the attacker correlating the HW of any specific sequence containing a target operation with the measurement traces.
Here, we instead consider that a potential attacker can leverage leakage information from different sequences containing the target operation at the same point. 
That is, if we executed $k-1$ MAC operations (out of the original $j-1$ operations) before the $j${th} original MAC, the $k$th leakage time point contains leakage of the $j$th weight with different sequences, each consisting of $k-1$ MACs plus the the $j${th} original MAC. 
Note that, such an adaptive attack is only theoretical since it is very challenging for an attacker to combine the multiple attack models for different MAC sequences at the same time point.

\subsection{Theoretical Leakage Upper Bound}
In this section, we present a theoretical upper bound for the leakage, demonstrating the effectiveness of \method even against a highly capable adaptive attacker. The derivation of the maximum proportion of leaked traces is detailed in Appendix~\ref{appendix: leakage upperbound} and given by:
\begin{equation}\label{eq:P.adapt}
    P = p^j \left(\begin{array}{c} j-1 \\ \lfloor pj \rfloor \end{array} \right) \left(\frac{1-p}{p}\right)^{j-\lfloor pj \rfloor -1}, \qquad \mbox{for } p<1.
\end{equation}
Here $\lfloor x \rfloor$ is the floor operation, i.e., the largest integer $\le x$.

\subsection{Mitigation Strength Against Adaptive Attack}
\begin{figure}[t]
    \centering
    \begin{subfigure}[b]{0.49\linewidth}
    \centering
        \includegraphics[width=0.99\linewidth]{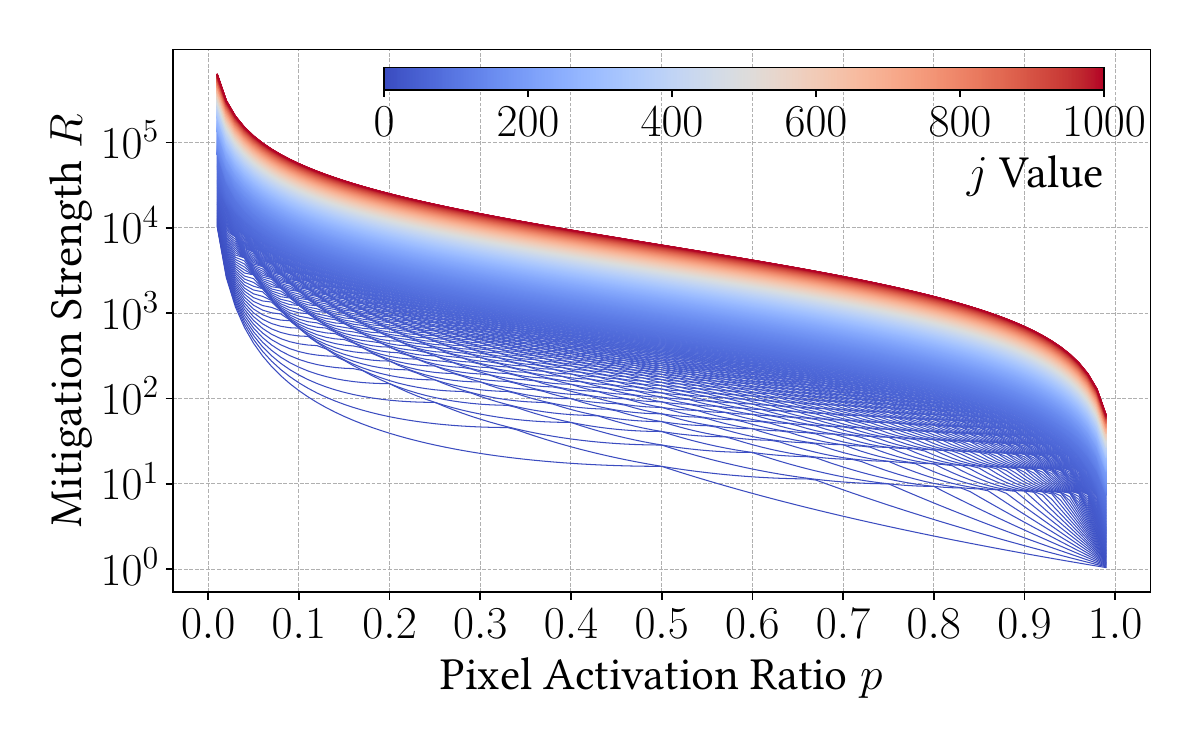}
        \caption{Visualize Lower Bound of Mitigation Strength $R$ under different $j$ for potential adaptive attack.}
        \label{fig: R-adaptive}
    \end{subfigure}
    \hfill 
    \begin{subfigure}[b]{0.49\linewidth}
        \centering
                   
                  \includegraphics[width=0.99\linewidth]{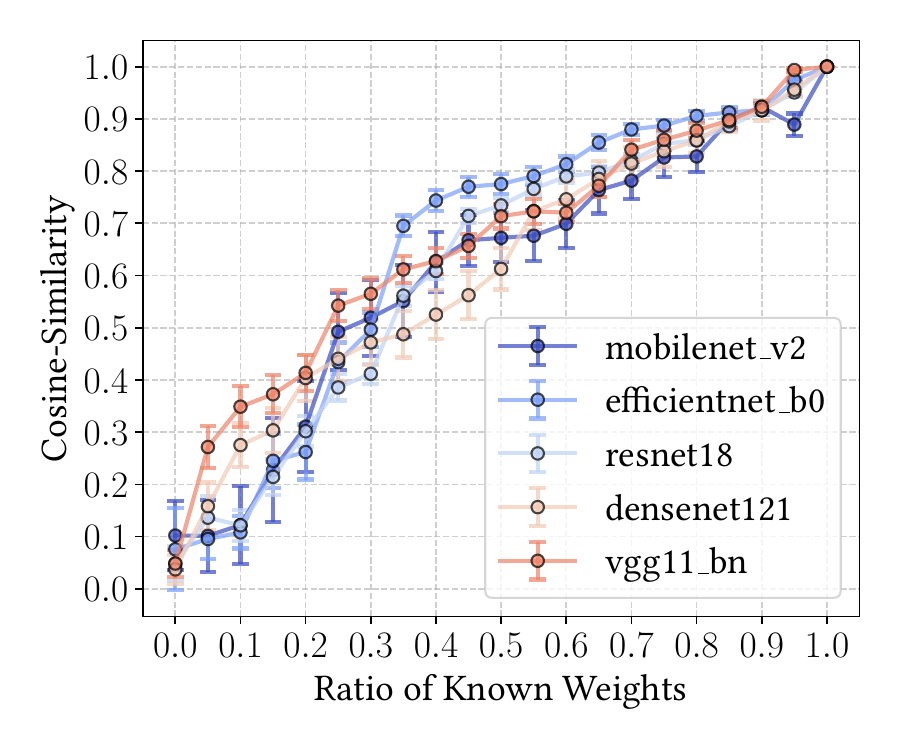}
        \caption{Attack Performance in Embedding Similarity}
        \label{fig: adaptive attack performance}
    \end{subfigure}
    \caption{Experiments on Potential Adaptive Attack}
    \label{fig: ablation study on alpha}
\end{figure}

Based on the previous analysis of the maximum leakage, the mitigation strength of \method against any potential adaptive attack is quantified as:
\begin{equation}
    \label{eq: R2}
    R=\left[\left(\begin{array}{c} j-1 \\ \lfloor pj \rfloor \end{array} \right) p^{\lfloor pj \rfloor +1} (1-p)^{j-\lfloor pj \rfloor -1} \right]^{-2}
\end{equation}
The theoretical adaptive attack's mitigation strength \( R \) is visualized in Fig.~\ref{fig: R-adaptive}. Table~\ref{tab: adaptive j} demonstrates that the same criterion \( j^* \), as shown in Table~\ref{tab: Nr vs p}, achieves a mitigation strength greater than \( 10^3 \) for different values of \( p \).
For instance, with \( p = 0.5 \), \( j \geq 160 \), which indicates that \method still effectively protects the majority of weights in a DNN model. For example, the first layer of MobileNet-v2 contains \( 864 \) weights. We further analyze the end-result of partial weight recovery on the model performance%
in Fig.~\ref{fig: adaptive attack performance}.
Specifically, for a pre-trained model, we compare the cosine similarity between the original embeddings and the embeddings from a model with partially known weights (with the remaining weights randomly initialized). For embeddings to retain semantic meaning, a cosine similarity above \( 0.8 \) is required, meaning the attacker must know approximately \( 70\% \) of the weights to replicate the victim model's functionality.
As shown in Table~\ref{tab: adaptive j}, for a victim model MobileNet-v2 and with \( p = 0.5 \) as the pruning ratio, only the first \( 18.6\% \) of weights can be stolen with $R\leq 1000$. This results in embeddings with a cosine similarity of 0.2, which lack practical semantic significance for the layer's functional representation.

\begin{table}[h]
  \centering
  \small
    \caption{Leaky time point $j^*$ for potential adaptive attacks}
  \begin{tabular}{c|c|c|c|c|c|c|c|c|c}
\hline
    $p$   & 0.1 & 0.2 & 0.3 & 0.4 & 0.5 & 0.6 & 0.7 & 0.8 & 0.9 \\
    \hline
    $j^*$ & 19   & 40   & 71   & 107   & 160   & 249   & 399  & 679  & 1610  \\
    \hline
  \end{tabular}

  \label{tab: adaptive j}
\end{table}

\section{Discussions and Future Works} \label{sec: conclusion}
In this section, we provide more insights about our proposed \method, including its limitations and potential future works.

\noindent\textbf{Applicable Devices:}
Our proposed method focuses on defending against SCA attacks on MCUs, which are the most vulnerable edge devices and are commonly targeted by attackers. The intrinsic parallel computation on more complex edge platforms poses challenges for SCA, and more advanced SCAs are required. Our proposed method still has the potential to protect complex systems against advnaced SCAs %
through careful analysis of the power model and effective leakage hiding.

\noindent\textbf{Scope of DNN models:}
In this work, we focused on image classification tasks using the MNIST and CIFAR-10 datasets on the MCU platform.
Our future work can explore more complex tasks such as object detection, image generation, or language models, which are usually deployed on more powerful platforms. Adapting our method to these advanced tasks and platforms will require further research and development.

\section{Conclusions} \label{conclusions}
This paper presents \method, a novel lightweight defense mechanism that counters DNN parameter retrieval based on side-channel analysis. It judiciously alters model inputs to thwart sequential SCAs targeting sensitive weights while preserving the model accuracy. This study underscores the effectiveness of \method and the importance of leveraging DNN characteristics for more efficient defense mechanisms.

\bibliographystyle{IEEEtran}
\bibliography{sample-base}
\appendix
\section{Derivation of Leakage Upper Bound for Adaptive Attack} \label{appendix: leakage upperbound}
When the original $j$-th operation occurs in the $k$-th position, there are $\left(\begin{array}{c} j-1 \\ k-1 \end{array} \right)$ possible such sequences of MAC operations.
These are all the possible sequences for the leakage related to the original $j$-th operation to occur on the $k$-th leakage time point. The proportion of traces corresponding to these sequences among all traces can be expressed as:
$$
\left(\begin{array}{c} j-1 \\ k-1 \end{array} \right) p^k (1-p)^{j-k}.
$$
If a potential adaptive attack can use leakage at the $k$-th time point from all these sequences together, the above expression quantifies the usable leakage as proportion of the original leakage for $j$-th operation without our countermeasure.
Hence, similar to the mitigation strength analysis in \ref{sec: theoretical R}, the largest leakage proportion $L_{max}$ over all time points is given by:
\begin{align*}
 L_{max}  &=  \max_{1 \le k \le j} \left(\begin{array}{c} j-1 \\ k-1 \end{array} \right) p^k (1-p)^{j-k} 
\\&=   p^j \max_{1 \le k \le j}
\left[\left(\begin{array}{c} j-1 \\ k-1 \end{array} \right) \left(\frac{1-p}{p}\right)^{j-k}\right]
\end{align*}

To find the expression of the maximum leakage point, we focus the change between $k$ and $k+1$. Specifically,

{\small
\begin{align*}
    L_{k} & = \left(\begin{array}{c} j-1 \\ k-1 \end{array} \right) \left(\frac{1-p}{p}\right)^{j-k} 
 = \frac{(j-1)!}{(k-1)!(j-k)!}\left(\frac{1-p}{p}\right)^{j-k} 
\end{align*}
\begin{align*}
    L_{k+1} & = \left(\begin{array}{c} j-1 \\ k \end{array} \right) \left(\frac{1-p}{p}\right)^{j-k-1}
    = \frac{(j-1)!}{(k)!(j-k-1)!}\left(\frac{1-p}{p}\right)^{j-k-1}
\end{align*}
}

The ratio between $L_{k+1}$ and $L_{k}$ is:
\begin{align*}
\frac{L_{k+1}}{L_{k}}=\frac{j-k}{k}\left(\frac{p}{1-p}\right)    
\end{align*}
This ratio is greater or equal than 1 when 
\begin{align*}
(j-k)p \ge k(1-p), \mbox{ or } \ \ jp \ge k.    
\end{align*}

Therefore, the maximum value of $L_{k}$ is achieved at
\begin{align*}
    k = \lfloor pj \rfloor + 1, \qquad \mbox{for } p<1.
\end{align*}

Thus, the largest leakage proportion is 
\begin{align*}
    p^j \left(\begin{array}{c} j-1 \\ \lfloor pj \rfloor \end{array} \right) \left(\frac{1-p}{p}\right)^{j-\lfloor pj \rfloor -1}, \qquad \mbox{for } p<1
\end{align*}
as in~\eqref{eq:P.adapt}. Note that this equals to 
$$
\left(\begin{array}{c} j-1 \\ \lfloor pj \rfloor \end{array} \right) p^{\lfloor pj \rfloor+1} (1-p)^{j-\lfloor pj \rfloor -1}.
$$
The mitigation strength is the inverse of this quantity squared, thus we arrive at the formula~\eqref{eq: R2}.

\end{document}